\documentclass[times,twocolumn,final]{elsarticle}
\usepackage[dvipsnames]{xcolor}

\usepackage{cag}
\usepackage{framed,multirow}

\usepackage{amssymb}
\usepackage{latexsym}

\usepackage[hyphens]{url} 

\definecolor{newcolor}{rgb}{.8,.349,.1}
\definecolor{plum}{HTML}{A02B93}
\definecolor{darkplum}{HTML}{50164A}
\definecolor{darkblue}{HTML}{084F6A}

\usepackage{hyperref}

\usepackage[switch,pagewise]{lineno} 

\journal{Computers \& Graphics}

\begin{document}

\verso{Preprint Submitted for review}

\begin{frontmatter}

\title{Visualization of Age Distributions as Elements of Medical Data-Stories}%
\author[1]{Sophia \snm{Dowlatabadi}\corref{cor1}}
\cortext[cor1]{Corresponding author: 
Tel.: +49-391 67-58772;  
fax: +49-391-67-41164;}
\emailauthor{sophia.dowlatabadi@st.ovgu.de}{Sophia Dowlatabadi}

\author[1]{Bernhard \snm{Preim}}
\author[1]{Monique \snm{Meuschke}} 

\address[1]{Department of Simulation and Graphics, University of Magdeburg, Universit\"atsplatz 2, 39106 Magdeburg, Germany}

    


\received{\today}

\begin{abstract}
In various fields, including medicine, age distributions are crucial. Despite widespread media coverage of health topics, there remains a need to enhance health communication. Narrative medical visualization is promising for improving information comprehension and retention. This study explores the most effective ways to present age distributions of diseases through narrative visualizations. We conducted a thorough analysis of existing visualizations, held workshops with a broad audience, and reviewed relevant literature. From this, we identified design choices focusing on comprehension, aesthetics, engagement, and memorability. We specifically tested three pictogram variants—pictograms as bars, stacked pictograms, and annotations. After evaluating 18 visualizations with 72 participants and three expert reviews, we determined that annotations were most effective for comprehension and aesthetics. However, traditional bar charts were preferred for engagement, and other variants were more memorable. The study provides a set of design recommendations based on these insights.
\end{abstract}


\end{frontmatter}



\section{Introduction}
\label{sec:intro}

In various disciplines, demographic data such as age is crucial, serving key roles from socioeconomic analyses to informing public health policies through statistical media reports. Understanding age-related data is particularly vital in healthcare, as evidenced by the German Centre for Cancer Registry Data at the Robert Koch Institute, which uses diverse diagrams to illustrate age distributions of cancer types~\cite{RKI_Hodenkrebs}. With age-dependent risk factors being critical in health communication and approximately 59\,\% of the German population exhibiting low health literacy~\cite{Schaeffer_2021}, there is a need to enhance health communication.

 Medical information presented in a narrative format is more memorable~\cite{Mittenentzwei_ArXive_2022, Meuschke_2022}. Known as narrative visualization, this technique combines narratives with visual elements to captivate and hold the audience's attention effectively~\cite{SegelandHeer_Narrative}. By integrating age distributions into narrative visualizations, we can enhance public understanding of disease risk groups. This study explores the best ways to display age distributions of diseases through narrative visualizations to improve public comprehension and engagement with health information.


In our study, we address the fundamental research question (\textbf{RQ}): \textit{How should the age distribution of a disease be presented in narrative medical visualization?} Using an adapted Double Diamond process~\cite[pp. 24-25]{DouDiam}, we analyzed a wide array of visualizations and pertinent literature to understand the current design space for age distribution visualizations. We also conducted a qualitative pre-study workshop with five participants to discuss a representative subset of these visualizations. This discussion led to the identification of 40 Design Choices (DCs) that enhance age visualizations, particularly focusing on the use of pictograms, one of the key DCs identified. 

We developed 18 visualizations featuring pictograms for case studies on breast cancer, salmonellosis, and bipolar disorder. These were qualitatively evaluated by three medical illustration experts and quantitatively tested by 72 participants, providing a broad spectrum of insights. From this analysis, we derived recommendations for future age distribution visualizations in medicine. Additionally, based on the fundamental RQ, we formulated three sub-questions to further direct our ongoing research in this crucial field.

\begin{itemize} 
\item \textbf{RQ.1} How are different visualization techniques and their respective chart elements currently utilized to display age in a medical context?
\item \textbf{RQ.2} How can different chart elements be added and modified to communicate the displayed age in narrative medical visualization?
\item \textbf{RQ.3} What are appropriate visualizations of age in narrative medical visualization regarding comprehension, memorability, aesthetics, and engagement based on the opinion of the broad audience as well as experts?
\end{itemize}

In summary, our contributions are:
\begin{itemize} 
\item We analyzed the visualization of age distributions (in the following short: age visualizations) from various contexts and discussed a subset in workshops with five broad audience participants. 
\item We evaluated age visualizations qualitatively with domain experts and tested them with a broad audience.
\item We derived design suggestions for narrative visualizations of age distribution. 
\end{itemize}

\section{Related Work}
In the following section, we discuss research on (medical) narrative visualization and current research on the perception of charts in their context. These studies provide a foundation for the current research and highlight its relevance.

\subsection{Narrative Medical Visualization} 

Segel and Heer opened up a new field of research when they labeled the combination of narratives and interactive graphics as narrative visualization – "visualizations intended to convey stories"~\cite{SegelandHeer_Narrative}, to engage broad audiences in their learning experiences~\cite{Kleinau}. Mittenentzwei et al.~\cite{Mittenentzwei_ArXive_2022} highlighted its potential in public health, empowering individuals of the general public to understand their health. 

The general public is diverse in terms of "knowledge and interests, […] age groups, […] cultural, geographical, and educational backgrounds.”~\cite{Boettinger_2020}. In the context of medical information, this group is comprised of medically interested audience, (potential) patients as well as relatives of those~\cite{Meuschke_2022}. The concept of public health plays a key role here: it is "the science and the art of preventing disease, prolonging life and promoting physical health and efficiency through [...] the education of the individual in principles of personal hygiene, the organization of medical and nursing service for the early diagnosis and preventive treatment of disease […].”~\cite{Winslow}. However, educating patients differs from educating the general public. Patients focus on treatments and predictions, while the public leans towards preventative measures.

Research indicates that narrative-embedded medical information is more memorable. Meuschke et al.~\cite{Meuschke_2022} detailed how medical visualization can narrate data-driven disease stories, proposing a seven-stage template including disease definition, treatment, and prevention. 
Mittenentzwei et al.~\cite{Mittenentzwei_ArXive_2022} illustrated targeting narrative visualization for cerebral small vessel disease, distinguishing between data-driven, context-driven, character-driven, and interactive content, using a story structure inspired by Campbell’s Hero’s Journey~\cite{Campbell2008hero}. 
Kleinau et al.~\cite{Kleinau} developed a data-driven narrative on aortic blood flow to educate about the rising prevalence of cardiovascular diseases.

These works mark initial attempts at applying narrative visualization to medicine, laying the groundwork for future endeavors. They contextualize age distribution visualizations, suggesting their inclusion into disease narratives to aid public understanding of certain diseases. Age, considered a medical risk factor~\cite{Niccoli2012}, could feature in the "Disease Prevention”~\cite{Meuschke_2022} stage or "Risk Factors”~\cite{Mittenentzwei_ArXive_2022} section of such narratives, prompting audiences to adapt their lifestyles.

\subsection{Information Visualization and Chart Perception in Narrative Contexts} 
%

Age visualizations, essential to narrative medical visualizations, provide insights into disease contexts, enhancing user understanding and revealing medically relevant insights. Within this framework, Hullman et al.~\cite{Hullman2011} introduced "visualization rhetoric," which outlines how design techniques influence interpretation in narrative visualizations. They delineate five rhetoric categories—information access, provenance, mapping, procedural, and linguistic-based—that impact viewer interpretation differently. Techniques like information reduction and filtering fall under information access, aimed at streamlining data presented to viewers. These categories span four editorial layers: data, visual representation, annotation, and interactivity. The base layer focuses on data display and manipulation, while the interactivity layer involves additional viewer interactions with the visualizations~\cite{Hullman2011}.

Furthermore, a large number of studies have already examined the perception of charts in the context of their associated components and how these can be modified. Studies, including those by Haroz et al.~\cite{Haroz_2015}, Bateman et al.~\cite{Bateman_2010}, Borkin et al.~\cite{Borkin_2013}, Skau et al.~\cite{Skau_2015} and Li et al.~\cite{Li2014}, explore different visual embellishments' impact on visualization perception. 
Haroz et al.~\cite{Haroz_2015} found pictographs enhance user engagement and recall but caution against their unnecessary use, which can distract users. They tested variations of how pictographs can be used. These included pictographs used in the x-axis, the bars themselves, or the background of the charts.
Bateman et al.~\cite{Bateman_2010} compare minimalist and complex visualizations, finding participants favoring more detailed visualizations for comprehension and memorability. Hereby, the authors tested a plain bar chart, such as Tufte~\cite{Tufte} advocates for, against a visualization created by the designer Holmes~\cite{Holmes_1984}.
Notably, several studies~\cite{Haroz_2015, Bateman_2010, Skau_2015} focus on bar charts, with some exploring visual features of charts that draw attention to certain data aspects~\cite{Kim_2021}, and others examining interaction in information visualization~\cite{Dimara_2020}. These works emphasize the importance of understanding how different design aspects influence visualization interpretation, user perception, and preference.

Our work aligns with these studies, investigating how various components and editorial layers affect the interpretation of age distribution visualizations. By examining these aspects against different criteria, our research provides insights for designing and evaluating age visualizations effectively.

\subsection{Criteria for Effective Visualizations} 
\label{sec:criteria}
\textit{Memorability} is an essential user experience goal~\cite{Saket_2016}. Many studies~\cite{Haroz_2015, Bateman_2010, Borkin2016, Kosara_C4PGV_2016} investigate the effects of embellishments on memorability, with contrasting views on their efficacy. E.g., Borkin et al.~\cite{Borkin_2013} argue against 'chart junk' while acknowledging, it may increase cognitive engagement and enhance data retention. Saket et al.~\cite{Saket_2016} also identify \textit{engagement} -"emotional, cognitive and behavioral connection [...], at any point in time [...], between a user and a resource.”~\cite{Attfield_2011}- as a crucial user experience goal. Mahyar et al.~\cite{mahyar2015} categorize engagement levels, with the highest being 'Decide', where users make decisions based on evaluations. In medical contexts, visualizations should not only prompt interaction but also empower users to make decisions relevant to their circumstances, fostering proactive health behaviors. Further, \textit{comprehensive} visualizations need to be functional, insightful, and enlightening~\cite[ch. 2]{cairo2016}, as they must accurately depict data and reveal nuanced insights, fostering audience acceptance~\cite{Kosara_C4PGV_2016,cairo2016}. Lastly, \textit{aesthetics} are still disregarded in the context of data and information visualization~\cite{Li_2020}. Through the aesthetic-usability-effect~\cite{UsabilityAesthEffect} as well as various studies~\cite{Cawthon_2007, Moere_2011, Filonik_2009} its importance for contribution to existing research is marked.
\section{Methodological Approach}
The subsequent section delineates the methodological approach employed to answer \textbf{RQ.1} and \textbf{RQ.2}. With \textbf{RQ.1}, we aim to explore the current design space of age visualizations for gathering corresponding modification aspects. To answer \textbf{RQ.2}, we provided a detailed discussion on how pictograms as a chart element can be used for age visualizations of diseases.

\subsection{Diseases for Context} 
The disease selection prioritizes a broad audience appeal. The World Health Organization (WHO) categorizes diseases into communicable and noncommunicable ~\cite{CommunNonCommunMental}. Mental health disorders, highlighted by WHO as affecting one in eight globally in 2022 ~\cite{MentalDisorders}, are also included, given their increasing relevance and global impact. Based on the disease distribution types and unique disease characteristics (e.g., number of new cases in recent years (incidence) or affected sex), the following three diseases were selected:

\paragraph{Noncommunicable Disease} Cancer is highly relevant to the daily life of a broad audience. Here, breast cancer, as one of the most common cancers, impacts both sexes but predominantly affects women, making it accessible for study participants. It also shows an age-related prevalence and a left-skewed distribution ~\cite[p. 23]{Sibbertsen2015}.
\paragraph{Communicable Disease} WHO reports annually one in ten foodborne disease occurences~\cite{Salmonella}. Due to its preventability and non-fatality, salmonellosis can positively engage study participants. Its right-skewed distribution~\cite[p. 23]{Sibbertsen2015} and impact on both sexes offer distinct features.
\paragraph{Mental health Disorder} Roughly 2\,\% of people worldwide have bipolar disorder (BPD)~\cite{BPD}. Stigma surrounds this disorder due to limited understanding~\cite{Hawke2013}. Also, its age distribution shows distinct peaks: late teens to early adulthood (18-24) and mid-30s to mid-40s~\cite{Manchia2017}.

\subsection{Criteria for Visualization Selection} 
\label{subsec:visselcrit}
The criteria presented in Section~\ref{sec:criteria} were applied as quality and evaluation criteria. Further, selection criteria are vital for ensuring that the visuals researched contribute effectively to the Research phase. These criteria include
\begin{itemize}
    \item prioritizing age information and data-driven visuals,
    \item including at least one chart and limiting other visual types, 
    \item addressing multiple age groups, and
    \item excluding map visualizations, 3D visuals, additional information coding.
\end{itemize}

\subsection{Adapted Double Diamond Process} 
\begin{figure}[t]
 {\includegraphics[width=1.0\columnwidth]{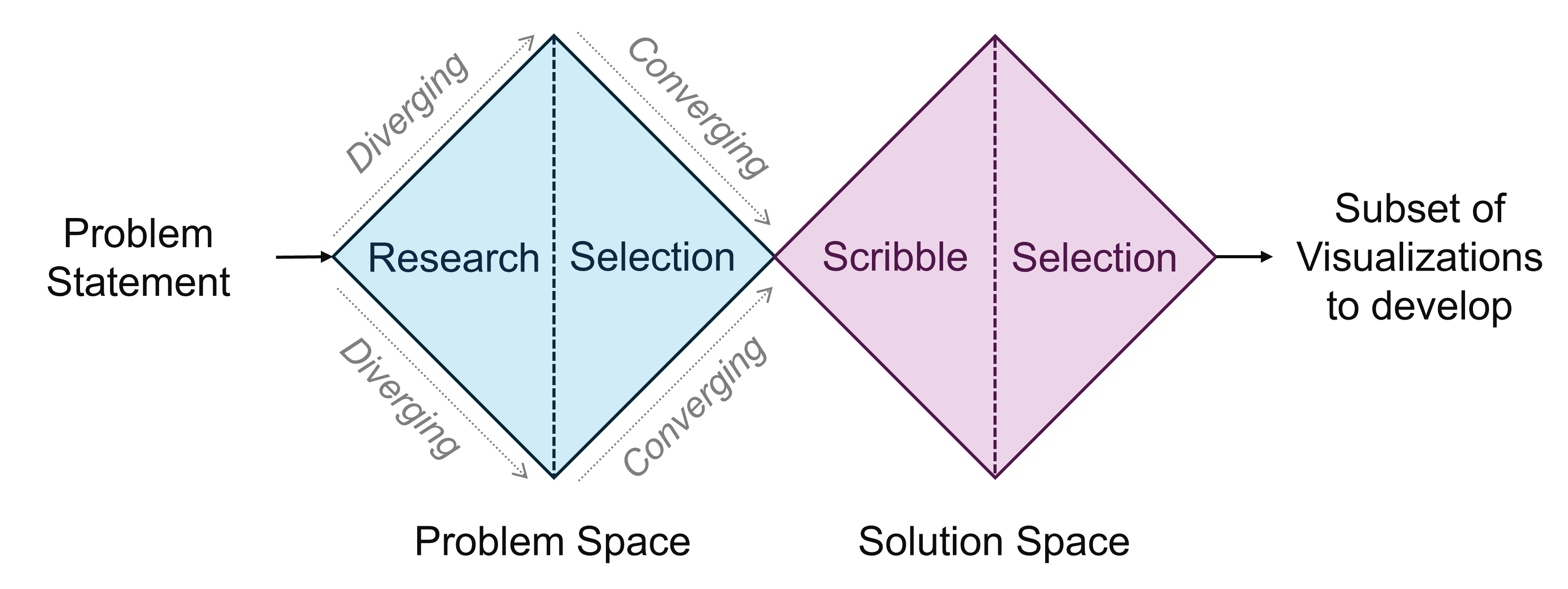}}
 \caption{Modified form of the Double Diamond process ~\cite{DouDiam} to visualize the methodological approach.}
 \label{fig:DoubleDiamond}
\end{figure}

Given the work's purpose of "presentation", the visualizations should aim at "insight" and "awareness"~\cite{Kosara_C4PGV_2016} in the broad audience. Thus, this work follows an adapted \textit{Double Diamond approach} ~\cite[pp. 24-25]{DouDiam} (see Figure ~\ref{fig:DoubleDiamond}) tailored to them as the target audience. This involves exploring and understanding problems, focusing on important issues, and generating high-level concepts. In every Diamond, the process \textit{diverges} to explore the research area widely and then \textit{converges} to evaluate solutions ~\cite[pp. 24-25]{DouDiam}. The process is initiated with a \textit{problem statement} (modified \textbf{RQ.1}) to set the goal of determining eligible chart types, annotations, and interactions for age visualization. The process ends with a \textit{subset of visualizations} to be implemented and tested. To achieve this, the four phases of the \textit{Problem} and \textit{Solution Space} diamonds are outlined in the following.

\begin{figure}[t]
 {\includegraphics[width=0.75\columnwidth]{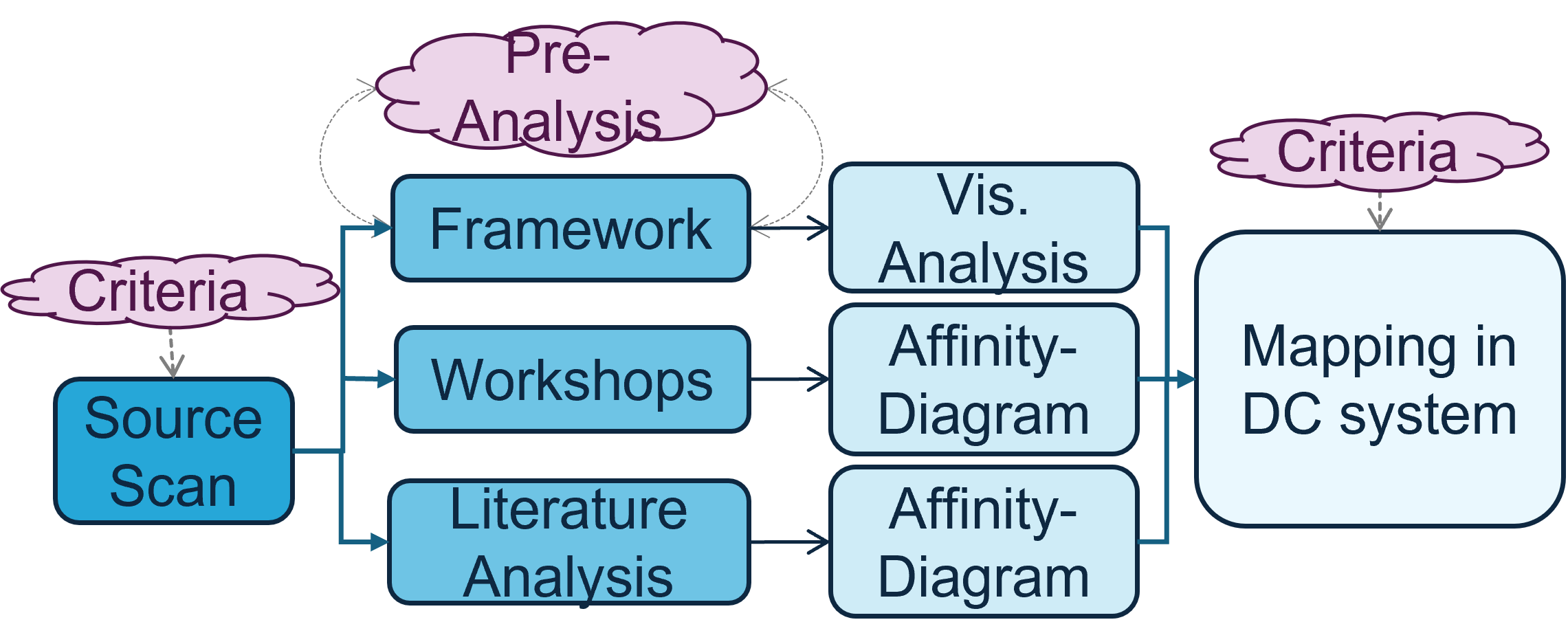}}
 \centering
 \caption{Process of the Research phase in the Problem Space.}
 \label{fig:Research}
\end{figure}

\subsubsection{Research}
The Research phase (see Figure~\ref{fig:Research}) was initiated with a \textit{source scan}. This involved researching existing visualizations from various sources (medical information pages and articles, online articles from diverse contexts, blogs and other freely available content, and scientific research studies). Here, the criteria mentioned in Section~\ref{subsec:visselcrit} were consulted. In the next step, \textit{literature} was analyzed, a \textit{workshop} with five participants from the broad audience was conducted and a \textit{framework} of chart components based on influential and fundamental literature was created. This was optimized through a pre-analysis of the found visualizations. After that, the 24 visualizations found were analyzed with the help of it, and the results of the workshops and literature analysis were organized through affinity diagramming~\cite[p. 12]{martin2013designmethoden}. Lastly, the results were mapped via consulting the criteria into a \textit{Design Choice (DC)} system. The DCs represent the modification aspects applicable to age visualizations. Through this system of DCs \textbf{RQ.1} is answered extensively with the focus on bar charts.

\paragraph{Framework} The structure of the framework is composed based on the \textit{editorial layers} of Hullman et al. ~\cite{Hullman2011} (data, visual representation, annotation, interactivity) with added concepts from other studies: In the \textit{data} layer the used metric, age bin widths and split in the represented data was analyzed. Chart element targets from Ren et al. ~\cite{Ren2017} were used in combination with visual properties described by Wilke ~\cite[ch. 2]{wilke2019fundamentals} and Tominski et al. ~\cite[p. 54]{tominski2020interactive} for the \textit{visual representation} layer. Ren et al.s'~\cite{Ren2017} work was also applied for the \textit{annotation} layer. Boy et al.'s ~\cite{Boy2015} \textit{interactivity} types were used for the last layer. Furthermore, \textit{narrative patterns} ~\cite{bach2018narrative} as well as \textit{rhetoric functioning}~\cite{Hullman2011} were regarded.

\paragraph{Workshop} The participants were diverse in terms of their ages (mean age: 37) and current live situations (interns, students, and employees). We conducted workshops comprising information, introductory, main, and closing phases, inspired by Misoch ~\cite [p. 68]{misoch2019qualitative}. In the \textit{information phase}, the context of the workshop is established by briefly explaining the background and topic. Information about the participants was obtained in the \textit{introduction phase}. The \textit{main phase} employed a modified 'triad method'~\cite[p. 186]{martin2013designmethoden}, where participants selected three visualizations as stimuli (a triad) from seven curated as per the source scan. After selection, they identified similarities and differences to gather qualitative insights, aiding problem space exploration~\cite[p. 186]{martin2013designmethoden}. Participants reviewed age pyramids, bar charts, line charts, and pictogram-based visuals from medical and other contexts. 
In the \textit{Closing} phase, the participants were asked for general feedback on the workshop and the method. 

\begin{figure}[t]
 {\includegraphics[width=0.7\columnwidth]{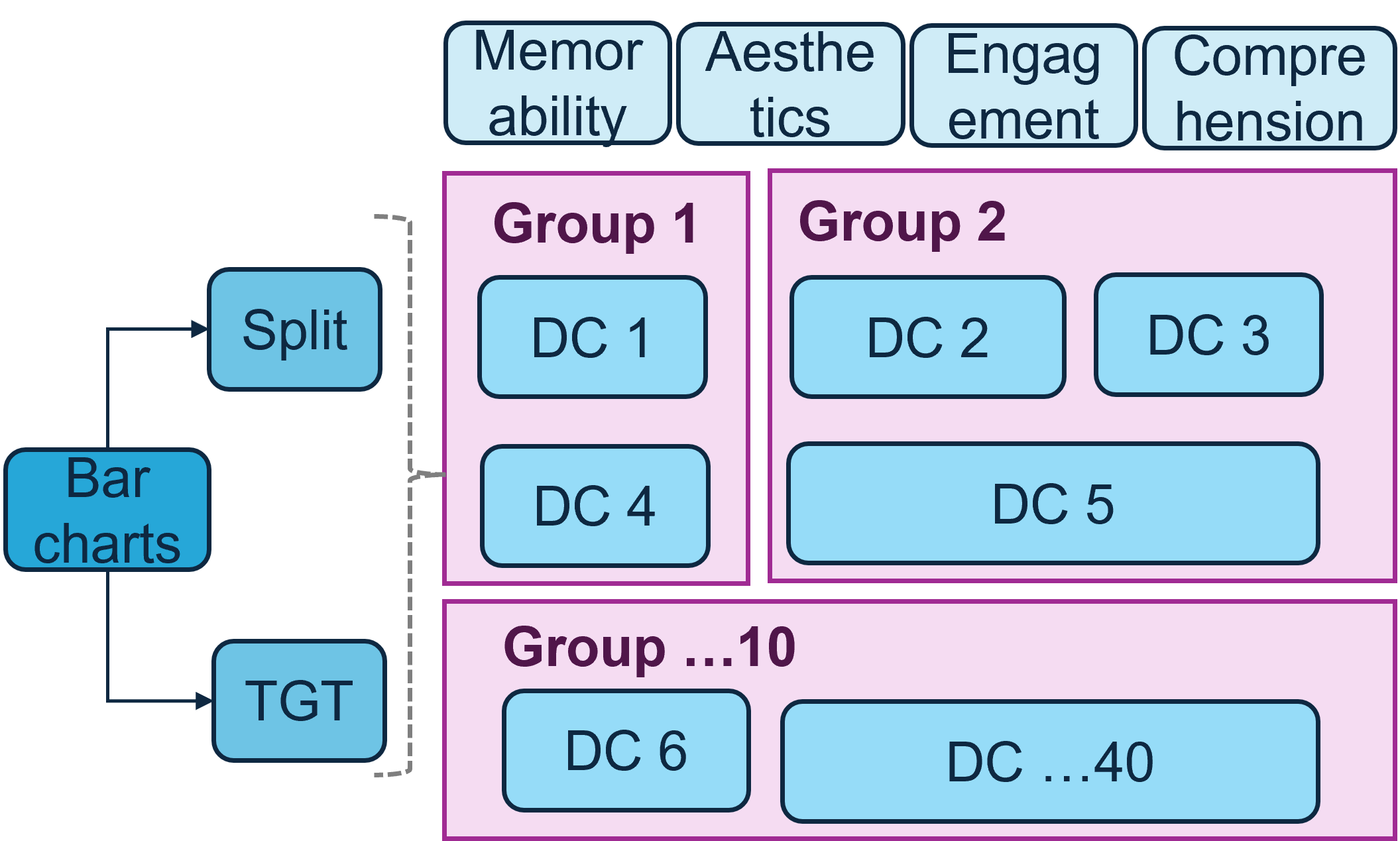}}
 \centering
 \caption{Illustrative overview of the DC system with the different dimensions.}
 \label{fig:DCsys}
\end{figure}

\paragraph{DC System}  The DC system (see Figure~\ref{fig:DCsys}) consists of five dimensions: The first dimension addresses the \textit{visualization technique} focusing on bar charts as the most prevalent in the analysis. The second dimension examines \textit{data splits}, crucial for analyzing both together (TGT) and separate (SPLIT) sex data. The third dimension encompasses the \textit{quality criteria}. \textit{DCs} are then aligned with these, offering multiple variations (fourth dimension), and are \textit{grouped into various categories} (fifth dimension): In total \textit{40 DCs} have been gathered regarding chart elements, annotations, interaction, animation, style, structure, information, age bins, data and other aspects of age visualizations.

\subsubsection{Selection}
In phase two, the focus is on selecting the most promising ideas. Among the many, one influential DC—\textit{pictograms}—is chosen to proceed as an example. Hereby, the analysis showed three Variants for utilizing pictograms: \vspace{3mm}\\  
\textbf{DC Variant A:} pictograms as bars\\
\textbf{DC Variant B:} stacked pictograms\\
\textbf{DC Variant C:} pictograms as annotations\\

\subsubsection{Scribble}
In the following, we explore pictogram utilization to answer \textbf{RQ.2}. Firstly, to diverge in the Solution Space, different considerations have to be made regarding color, metric, age bin width/pictogram utilization, and pictogram research. 
\begin{figure*}[t]
 \centering
 {\includegraphics[width=1.0\linewidth]{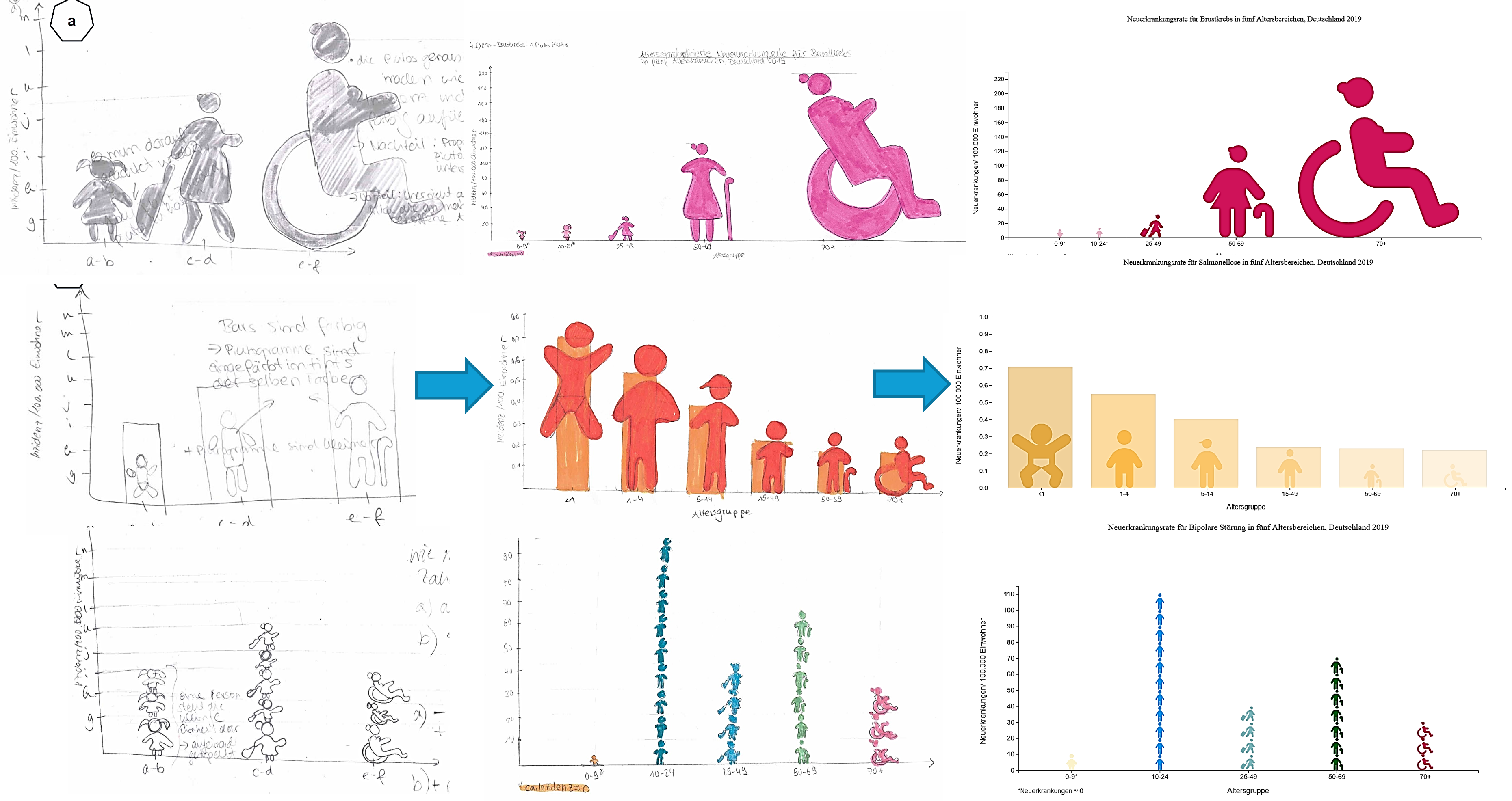}} \caption{Process of low-fidelity wireframes to final implementation of the three pursued ideas for breast cancer (first row), salmonellosis (second row) and BPD (third row). Left column: low-fidelity wireframes from the Scribble phase in the Solution Space. Middle Column: Wireframes with specific design aspects like color from the Selection phase in the Solution Space. Right column: Final visualizations implemented.}
 \label{fig:Sketches}
\end{figure*}

\paragraph{Basic Visualizations} Basic visualizations are composed of all analyzed properties that are not DC properties, i.e. that have contributed to the creation of DCs. These basic visualizations form the underpinning for the visualizations with pictograms.

\paragraph{Age bin width and pictogram research} There is one crucial decision regarding age bin width and pictogram research. Either many small age bins are used, allowing for higher accuracy at the expense of requiring numerous pictograms, or fewer and larger age bins are employed, which results in lower accuracy. The last option was selected as freely available pictograms from Fontawesome\footnote{https://fontawesome.com/icons/} and Flaticon\footnote{https://www.flaticon.com/} had to be relied on or had to be created by ourselves. Regarding the data sources (see Section~\ref{sec:sources}), the following age bins for the respective diseases were used: 
\vspace{3mm}\\ 
\textbf{Breast Cancer:}  0-9; 10-24; 25-49; 50-69; 70+ \\
\textbf{Salmonellosis:} \textless{1}; 1-4; 5-14; 15-49; 50-69; 70+\\
\textbf{BPD:} 0-9; 10-24; 25-49; 50-69; 70+

\paragraph{Pictograms} Pictograms represented age groups, associating each with relevant symbols like a baby for \textless{1}. They differ subtly for the male and female sex through implied secondary sexual characteristics (breasts), accessories (e.g., bow for the female baby or cap for the adolescent boy), or clothing (dress for women). Recent discussions in the design community~\cite{Dive_d} advocate for more inclusive and diverse pictograms, regarding factors like skin color and ethnicity. Limited options led to using pictograms but some choices, like the walking stick, may perpetuate negative stereotypes and reproduce ageism as social discrimination~\cite{Ageism}. Similarly, the use of the accessibility symbol for the 70+ age group, which is critically discussed~\cite{Domnguez2013}, may present an inaccurate view of inclusivity. While not intended to discriminate, such choices highlight the need for more inclusive pictograms.

\paragraph{Color} In the selection of colors, we considered the characteristics of the disease depicted and derived inspiration from the analysis of existing age visualizations.

\paragraph{Low-fidelity wireframes} Low-fidelity wireframes~\cite{Wireframing} (see Figure~\ref{fig:Sketches}) were created using pen and paper to explore various visualization options. This classic method allowed for idea generation without distraction regarding feasibility ~\cite{Roberts_2011}.

\subsubsection{Selection}
The following three ideas are selected for the DC Variants, converging the Solution Space. Wireframes (see Figure ~\ref{fig:Sketches}) are created with specific design elements like color. 
\vspace{3mm}\\ 
\textbf{DC Variant A:} The pictograms of the individual age groups represent the incidences of the diseases by their height. \\
\textbf{DC Variant B:} Incidences are rounded up, displaying pictograms as a whole. One is the smallest unit (representing an incidence of 1:100.000).\\
\textbf{DC Variant C:} The pictograms are centered in the bars of their respective age group. The bars are in focus through coloring the pictograms in a light tint, appearing 'transparent'.

\paragraph{Incidences close to zero} During the creation of the wireframes, handling pictograms with incidences close to zero was considered. Despite minimal occurrences, pictograms were retained to maintain consistency in age representation. Also, reducing size and color saturation were employed to signify low incidence.

\section{Implementation}
In this section, we describe the implementation of the visualizations (see Figure ~\ref{fig:Sketches}), which will be used for the evaluation::

\subsection{Data Sources}
\label{sec:sources}
For data on the selected diseases, two approaches were considered: using data from reputable institutions or generating synthetic data. While real data lends credibility, synthetic data can fill gaps when real data is inaccessible or incomplete ~\cite{Mannino2019}. Research revealed WHO collaborates with IHME, offering comprehensive health data through the Global Burden of Diseases (GBD) study ~\cite{WHOCollabIHME, GBD}. This data, available via an interactive website, was chosen due to the availability of all diseases and range of age bins ~\cite{GBD_Interactive}. 

\subsection{Structure of Datasets} 
The datasets were provided as CSV files. SPLIT includes separate entries for each sex within age groups. Key attributes include sex (categorical variable), age group (categorical), and incidence value (continuous). TGT follows a similar structure, listing incidences for age groups without sex differentiation.

\subsection{Software} 
The visualizations were implemented using the D3.js library\footnote{https://d3js.org/}, well-suited due to its ability to parse CSV data and bind it to visual elements. D3 facilitates creating pictogram-based visualizations in conjunction with HTML, CSS, and JavaScript.

\subsection{Adjustments to the Visualizations} 
\begin{figure}[t]
 {\includegraphics[width=1.0\columnwidth]{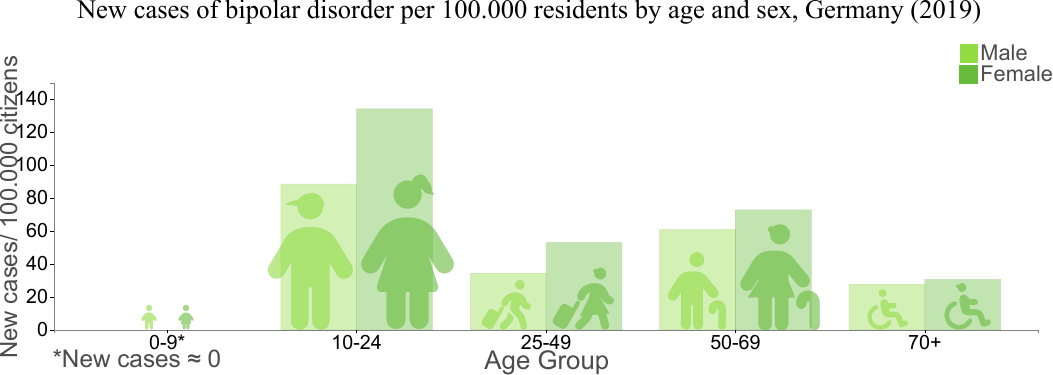}}
 \caption{SPLIT visualization of BPD for DC Variant C. It can be observed that for the age groups 10-24 and 50-69, the pictograms partially protrude.}
 \label{fig:418}
\end{figure}

During implementation, adjustments were made. One issue was a misalignment of pictograms in DC Variant C, especially in SPLIT visualizations, due to larger sizes with higher incidences, requiring protrusion from the bars (see Figure ~\ref{fig:418}).

\section{Evaluation}
To answer \textbf{RQ.3} expert reviews and a user study were conducted. Thus, the implemented visualizations utilizing pictograms can be evaluated. Both concepts will be outlined in the following:

\subsection{Expert Reviews} 
Expert reviews were conducted with three specialists who have between 5 to 20 years of experience in medical illustration and visualization. Their perspectives helped to identify visual design problems. Expert reviews involved semi-structured interviews, allowing experts to verbalize initial thoughts, impressions, and potential improvements using the Think Aloud method~\cite{ThinkingAloud}. The procedure, guided by Tory and Möller's recommendations (selecting experts carefully, conducting individual interviews, and maintaining neutrality)~\cite{Tory2005}, included an introduction, visualization presentation, and closing phase. The experts explored visualizations, provided feedback, and compared preferences. Experts reviewed six visualizations of BPD, both SPLIT and TGT versions of each DC Variant. 

\subsection{User Study} 
 Conducted as an online survey, participants were recruited from various networks (e.g., student communities and sports clubs). The study design balances between- and within-subjects approaches to accommodate the variables: DC Variants, disease types, and split dimension. One participant views either SPLIT or TGT visualizations to avoid data overload. Further, they observe all diseases and all DC Variants, yet a different DC Variant for each of the three diseases. The study is divided into six groups, with diseases shown in different sequences. The survey begins with an introduction, followed by questions on personal information and descriptions of the diseases. Tasks in the study focus on comprehension, aesthetics, engagement, and memorability with individual and group tasks.

 \paragraph{Comprehension} 
 Participants were asked about the highest incidence age group (understanding the basic intent of the visualization), reading specific incidences, and if they could deduce concrete assertions, drawing from Skau et al.~\cite{Skau_2015} and Mahyar et al.~\cite{mahyar2015}. They rated confidence and visualization suitability after each question. Then, they ranked visualizations for comprehensiveness as a group task.
 
 \paragraph{Aesthetics} 
 Participants rated their perceived 'beauty' of the currently observed visualization on a slider from "ugly" to "beautiful," following Cawthon and Moere~\cite{Cawthon_2007}. Additionally, they ranked visualizations as per aesthetic preference, allowing direct comparison~\cite{Cawthon_2007}.
 
 \paragraph{Engagement} 
 Participants rated statements after observing individual visualizations, assessing their inclination for further research of the depicted diseases, considering preventive measures, and informing acquaintances. This gauged the visualizations' impacts on encouraging action. The group task, adapted from Haroz et al.~\cite{Haroz_2015}, compared pictogram visuals to basic bar charts and text. Participants imagined explaining the depicted disease age distribution to another person and chose between the options. Based on Bateman et al.~\cite{Bateman_2010}, the participants explained their preference by selecting words from a list of positive words inspired by Garrison et al.~\cite{Garrison} (for the selected option) and negative words (for the non-selected options).  

 \paragraph{Memorability} 
To assess short-term memory performance, participants had to recall pictograms from the first visualization. They identified which pictogram represented a specific age group, ensuring relevance to the disease. Task complexity was enhanced by offering similar options and avoiding size differences among pictograms. Participants could also opt-out if unable to recall, maintaining data integrity. Further, participants had to recall visualization details, based on Borkin et al.~\cite{Borkin2016}.

\section{Results and Discussion}
This section presents the results of the expert reviews and the user study. 

\subsection{Participants}
A total of 72 participants (37 male, 34 female, 1 diverse) took part in the study. The majority of the participants were German (93\,\%). Most of the participants were between 26 and 35 (37,5\,\%) and between 18 and 25 (33\,\%). 

Most of the participants 62,5\,\% had a university degree. The lowest number of participants had an intermediate educational qualification with 4\,\%.
The majority of participants (61\,\%) did not have a visual impairment. How much the participants engaged with visualizations in their everyday lives was quite diverse, with 19 participants (26\,\%) rarely or sometimes, 15 (20\,\%) never, 12 (17\,\%) very often, and 2 (3\,\%) often. However, 31 of 72 participants (42\,\%) never engage with medicinal topics. Other than this, 16 participants (22\,\%) rarely and 13 (18\,\%) sometimes engage with medicinal topics.

\subsection{Expert Reviews and User Study} 
The results are comprised according to the four criteria comprehension, aesthetics, engagement, and memorability. 

 \paragraph{Comprehension} 
The experts found DC Variant C as the easiest for reading the data. This is due to the fact, that they are the most similar to conventional bar charts. Nevertheless, they also emphasized DC Variant A, the attention-grabbing property of the bigger pictograms, which helps to quickly grasp incidences. For DC Variant B, the stacked pictograms function as bars and, therefore, are also suited for that purpose. Moreover, the experts did highlight the importance of depicting the incidences for the biological sexes separately, as more information is available to the viewer. \\ 
The majority of participants were able to understand the basic intent of the visualization (DC Variant A: 96\,\%, B: 99\,\%, C: 82\,\%) and deduce insights (DC Variant A: 74\,\%, B: 71\,\%, C: 79\,\%) through all DC Variants. \\
But for DC Variant C, the majority of participants read exact incidences correctly (DC Variant A: 71\,\%, B: 75\,\%, C: 82\,\%) and were the least uncertain in their answers (DC Variant A: 22\,\%, B: 16\,\%, C: 5\,\%). This tendency can also be seen in the rankings, as DC Variant C was ranked first by 64\,\% compared to Variant A (17\,\%) and B (20\,\%)
Yet, the suitability for all DC Variants is rather low (DC Variant A: 46\,\%, B: 44\,\%, C: 45\,\%). The explanation for these results is two-fold.\\
First, regarding incidences close to zero, DC Variants A and B for both TGT and SPLIT had similar incorrect answers (DC Variant A TGT: 84\,\%, SPLIT: 79\,\%, B TGT: 100\,\%, SPLIT: 50\,\%), so they are not suited. This can be explained with the direct display of the incidences through the pictograms visually. In comparison, for DC Variant C the missing bar marks the incidence, resulting in only 31\,\% answering incorrectly. Hereby, the experts suggested either omitting age groups with incidences close to zero, using annotations directly at the pictogram, or enabling interactivity to show an annotation. \\
Second, the qualitative feedback of the participants for general legibility problems must be considered. For \textbf{DC Variant A}, there are two aspects criticized by the experts and study participants:
\begin{enumerate}
         \item 'Roundings' of the pictograms terminal point do not provide an appropriate reading point so the user study participants have difficulties at where to read the value from. Gridlines would have improved the experience.
         \item The proportions of the pictograms in terms of height and width make it very difficult to interpret the incidences correctly.
\end{enumerate} 
For \textbf{DC Variant B}, also two aspects are commented:
\begin{enumerate}
         \item It was distracting to count the individual pictograms. Here, the use of a data label was mentioned.
         \item The general form of depiction via the stacking was found distracting.
\end{enumerate}
The first point regarding DC Variant B referred to visualizations, where the y-axis was divided into smaller intervals. This affected the number of pictograms stacked. If the y-axis is divided into larger intervals, naturally, fewer pictograms have to be stacked. This was positively noted by the experts and regarded as helpful for the broad audience for legibility and perception of pictogram details. \\
For \textbf{DC Variant C}, the non-uniform sizes of the individual pictograms are confusing, as the participant did not know if the bars or the pictograms should have been regarded for reading. The experts proposed that the pictograms could be much smaller and share a uniform size as well as be placed at the same position for all age groups to avoid protrusion of bars.

Skau et al. ~\cite{Skau_2015} align with the perception that basic bar charts perform best compared to adaptations like rounded tops or triangles. They noted the importance of clear lines at bar ends for extending to the value axis. Furthermore, additional aids like gridlines and data labels, though helpful for clarity, should be used sparingly to avoid unnecessary segmentation ~\cite{Haroz_2015, Kosara_C4PGV_2016}.
 
\begin{figure}[t]
\centering
 {\includegraphics[width=0.85\columnwidth]{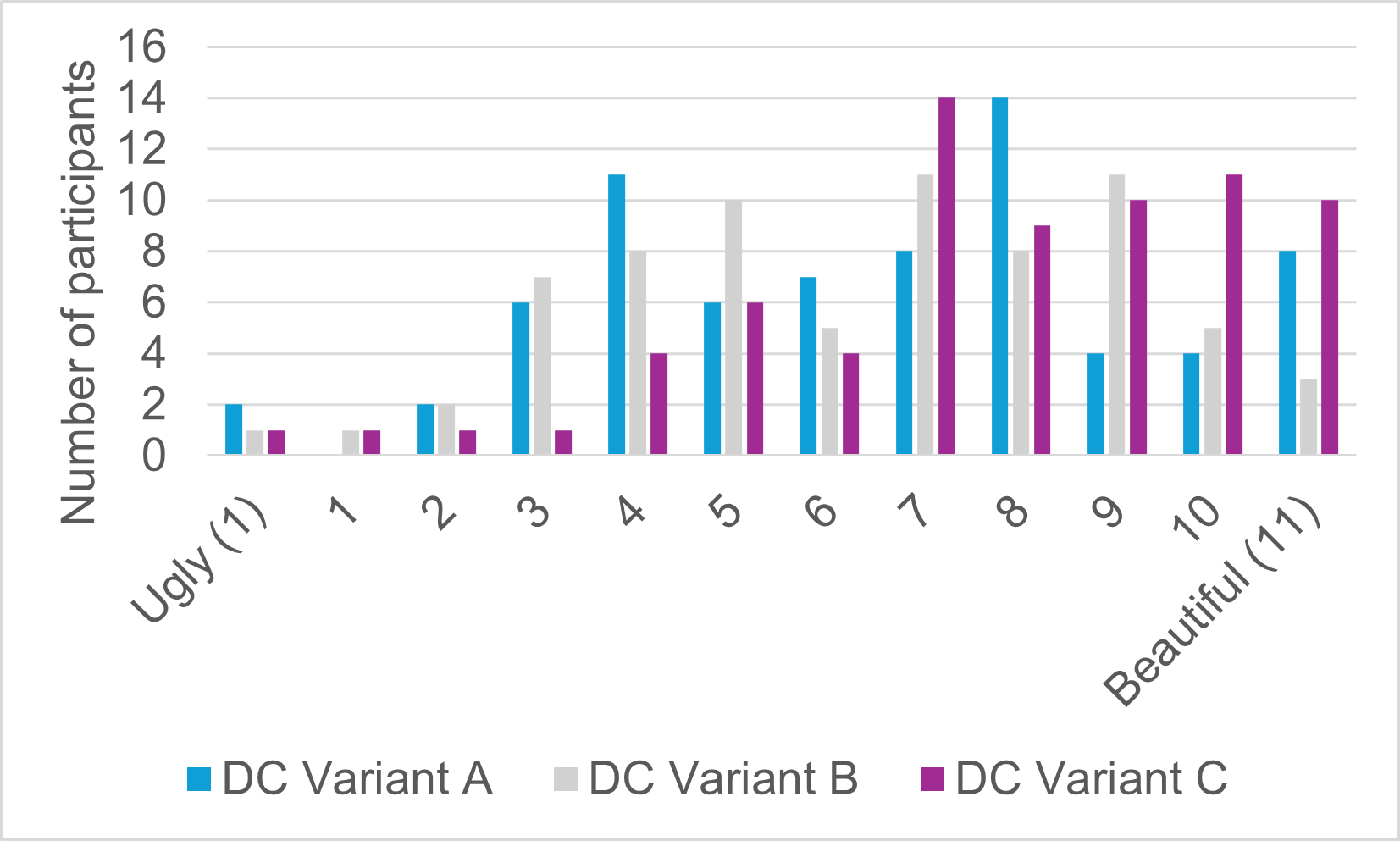}}
 \caption{Ratings of subjective aesthetics regarding the visualizations for DC Variant A, B, and C (N=72).}
 \label{fig:Aesth}
\end{figure}

 \paragraph{Aesthetics} 
Based on the results of the individual rankings (see Figure ~\ref{fig:Aesth}), DC Variant C was perceived as the most aesthetic.
This can also be seen in the results of the rankings. DC Variant C was placed first by 51\,\% of the participants, compared to 17\,\% Variant B and 32\,\% Variant A. However, the feedback indicates that rankings were influenced more by usability than aesthetics, aligning with the Aesthetic-Usability-Effect~\cite{UsabilityAesthEffect}, suggesting participants integrated their experiences into their ratings. Therefore, the results for aesthetics can not be viewed as unbiased.

Nevertheless, the experts and participants perceived the older age groups discriminatorily. Further, the depiction of sex was found stereotypical by the participants. According to the experts, \textit{either} designated pictograms for the individual sexes (even if stereotyping) should be utilized for ensuring accessibility \textit{or} androgynous pictograms should be used. Further, one expert expressed: \textit{"it is not about depicting age through the pictograms but showing people"}. By that, they mean that a) if the goal is to foster engagement through utilizing pictograms: the engagement with the visualizations would have been ensured even with depicting the same pictogram across different age groups, or b) if the goal is to foster engagement through being inclusive: per age group, a small set of pictograms could be used, that show different life situations.
 
\begin{figure*}[t]
 \centering
 {\includegraphics[width=1.0\linewidth]{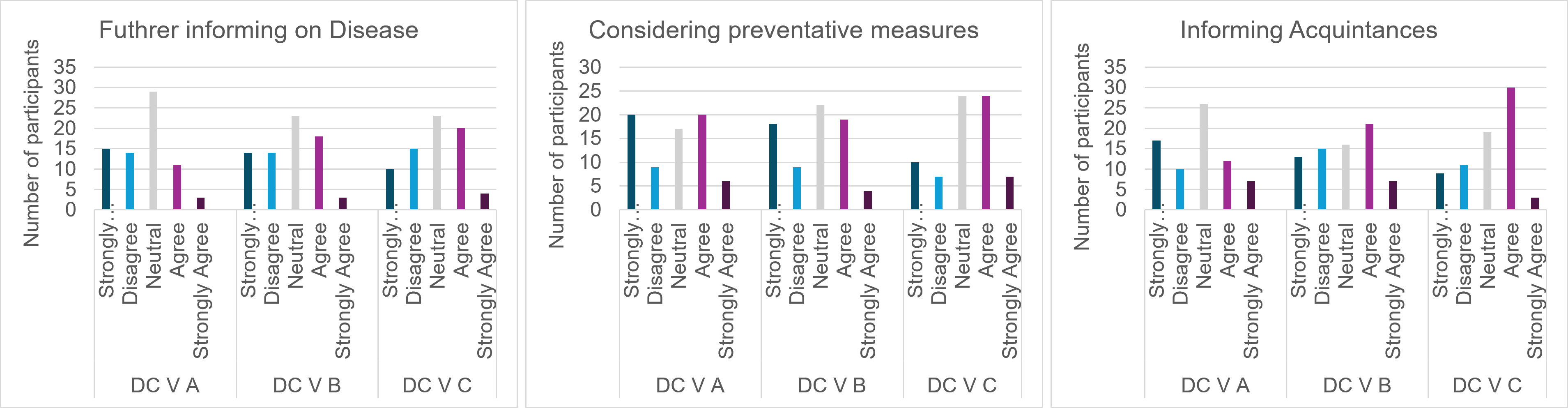}} \caption{Distribution regarding the ratings for the statements across the different DC Variants (DC V: DC Variant, \textcolor{darkblue}{Strongly Disgaree}, \textcolor{cyan}{Disagree}, \textcolor{lightgray}{Neutral}, \textcolor{plum}{Agree}, \textcolor{darkplum}{Strongly Agree}).}
 \label{fig:Engag}
\end{figure*}

 \paragraph{Engagement} 
As can be seen in Figure ~\ref{fig:Engag}, disagreement and neutrality are always present. Hereby, the participants often mentioned a lack of symptoms. This can be considered in the context of Mittenentzwei et al. ~\cite{Mittenentzwei_ArXive_2022} because the general public finds preventive measures for education the most interesting. Mittenentzwei et al.~\cite{Mittenentzwei_ArXive_2022} also expressed through the patients' disease journey that the patient has to perceive symptoms in their normal life to meet a physician. Therefore, this curiosity for symptoms can be credited to the fact, that the broad audience does not perceive the incidence of the risk groups as being an important enough aspect for engaging with a disease.

Besides that, if the participants are already aware of diseases, they tend to disagree with the statements. If they a) knew a little bit or b) were shocked about some aspect or c) the disease corresponded to their current life situations, they tended to agree with the statements. For example, for DC Variant A TGT one female participant in the age group 26-35 agreed with all statements for salmonellosis because she wants to grow a family in the near future. For DC Variant C SPLIT, a female (18-25) with BPD has previously undergone diagnostic procedures for mental disorders and advocates for the importance of seeking professional help.

Nevertheless, in all DC Variants, the majority of participants preferred the basic visualizations (DC Variant A: 74\,\%, B: 85\,\%, C: 63\,\%). Across the different DC Variants, the participants that selected the basic visualizations over the DC Variants gave the reasons (pro terms) of "easy to read/understand" (DC Variant A: 93\,\%, B: 95\,\%, C: 88\,\%) and "clear" (DC Variant A: 86\,\%, B: 86\,\%, C: 88\,\%) the most. The selection of the pro terms for the basic visualizations was similar to the user preferences in the study by Li et al.~\cite{Li2014}.
This can be attributed to a 'pure' utilization of pictograms: the visualizations have been implemented as if they were regular bars. Some participants described a general antipathy towards utilizing pictograms for 'professional charts' (as the visualizations were perceived by them), which also corresponds to Li et al.'s study~\cite{Li2014}. Experts also mentioned that certain aspects of the visualizations are not suitable for the broad audience: 
 \begin{enumerate}
         \item Y axes were used.
         \item A not 'fun' font and title was integrated. 
         \item No interaction was provided.
         \item The visualizations were not integrated into a story but brought into context through a scientifically oriented text. 
     \end{enumerate} 
Brinch states: "The viewer’s perception has both a cognitive and an emotional side, both of which are activated when encountering the data visualization and the context in which it is found [...]"~\cite{Brinch}. Thus, these factors could have affected 'passive' and 'active' engagement, as discussed by Bach et al.~\cite{bach2018narrative}. To enhance passive engagement, emphasis on emotions and empathy, possibly through rhetorical questions, could be beneficial. For active engagement, interactivity is key~\cite{bach2018narrative}. Various narrative patterns, like 'Make-a-guess'~\cite {bach2018narrative}, as seen in interactive charts by The New York Times ~\cite{NYT}, could stimulate audience curiosity. Exploring these strategies could improve engagement in the visualizations.\\
One aspect, that the experts highlighted positively, was the SPLIT visualizations in general: due to the sex-split depiction of incidences, more information could be gathered as \textit{"another element to the story is included."}

 \paragraph{Memorability} 
DC Variant A and B were the most memorable compared to DC Variant C, as 50\,\% in DC Variant A and 85\,\% in DC Variant B remembered the correct pictograms. For DC Variant C, 45\,\% remembered correctly. Hereby, Bateman et al. stated, “[…] the emotions, in combination with the visual imagery, help to anchor chart details in a viewer‘s memory." ~\cite{Bateman_2010}. As DC Variant A and B directly display the incidences visually through the pictograms, the emotional reactions (whether good or bad) of the participants with them, could have facilitated enhanced memorability.

\subsection{Design Suggestions} 
From the results, the following suggestions have been derived:\vspace{3mm}\\ 
\textbf{Design and Utilization of Pictograms}

\begin{enumerate}
    \item To \textit{depict the prevalence of certain age groups} for specific diseases, use \textit{pictograms as bars}.
    \begin{enumerate}
         \item \textit{Pictograms should not feature round or pointed terminal points} via heads and accessories, in order to maintain a clear contour for reading data.
     \end{enumerate} 
    \item Use \textit{stacked pictograms}, for \textit{memorability} and \textit{attention-grabbing}.
     \begin{enumerate}
         \item \textit{Avoid too many pictograms}, as the cognitive load of the viewer increases with the number of displayed pictograms. Alter the scaling for the y-axis to have larger incidence intervals.
         \item Pictograms should exhibit as \textit{little detail} as possible so that these are perceivable for smaller scaling formats.
     \end{enumerate} 
    \item For enabling \textit{viewer acceptance}, improved \textit{legibility} and overall \textit{higher engagement}, use \textit{pictograms as annotations}.
    \begin{enumerate}
         \item The \textit{pictograms should be uniform in size}, or else confusion can occur for the viewer as to regard the bars or the pictograms for reading the incidence.
         \item Regarding positioning in relation to the size of pictograms, two alternatives exist:
         \begin{enumerate}
             \item Make the bars a little wider/higher than the width/height of the pictograms, as overlapping and 'caged in' pictograms are criticized. Put the pictograms in the center of the bars.
             \item Scale the pictograms much smaller (but the same size) than the bars and place them in the same position for all age groups. This helps with layout clarity.
         \end{enumerate} 
     \end{enumerate} 
\end{enumerate} 

\textbf{Improve comprehension through readability}

\begin{enumerate}
    \item \textit{If you use a y-axis, employ gridlines}. The incidences of the higher age groups are more readable.
    \item \textit{If you do not use a y-axis, data labels should be exploited}. This has space-saving advantages as well as improves comprehension.
   \item For incidences close to zero, different approaches can be considered:
    \begin{enumerate}
         \item Omission: Omission of the age group with this incidence.
         \item Annotation: Write a remark regarding the incidence directly at the pictogram/age group as an annotation, and do not work with footnotes.
         \item Interaction: Display incidence as such and work with interaction options.
          \item Visual: Visually distinguish pictograms for different age groups to enhance clarity. Use this approach in conjunction with Annotations or Interactions for best results.
        \end{enumerate} 
\end{enumerate} 

\textbf{Improve acceptance through information}
\begin{enumerate}
\item \textit{Display the incidences for male and female split}, as these tend to be \textit{more informative} for the broad.
\item \textit{Depict the encoding of the data in a legend}, to explain how the data should be read.
\item \textit{Include disease symptoms} with age data in visualizations, as age alone does not sufficiently engage or prompt proactive responses.
\item Do not use pictograms that evoke negative associations. \textit{Use generic pictograms} instead. \textit{Address sex in SPLIT visualizations} to visually distinguish between male and female incidence rates and \textit{enhance accessibility}.
\end{enumerate} 

\subsection{Limitations} 
Several limitations apply to different aspects of this work:

\paragraph{\textit{Limited scope of visualization techniques}} In addition to bar charts, other visualization techniques, such as line charts, have been discovered. Accordingly, the system of DCs is limited to only one visualization technique and can be expanded in the future.
\paragraph{\textit{Pictogram variations}} The pictogram-based visualizations we implemented for the three DC Variants posit only a limited number of solutions. Future research should investigate alternatives for these and explore additional visualizations utilizing pictograms to enrich the analysis.
\paragraph{\textit{Limited number of participants}} The study's participant pool was too small to draw generalized and definitive conclusions. Future tests with larger participant groups are needed to establish correlations and significance more conclusively.

\section{Conclusion and Future Work}

This study has effectively demonstrated how age distributions of diseases in the context of medical data stories can be visualized, addressing three pivotal research questions. \\
RQ.1 and RQ.2 have been answered by successfully adapting the Double Diamond approach. RQ.1 was answered through extensive analysis of literature and visualizations from different contexts as well as the conduction of workshops. Thus, 40 DCs concerning modification aspects of bar charts were identified. For RQ.2, one of the most prominent DCs - pictograms - was explored by designing 18 visualizations regarding the utilization of pictograms as bars, stacked pictograms, and annotations. Lastly, RQ.3 was successfully resolved through reviews with domain experts and a user study with 72 participants from a broad audience.

Our analysis has highlighted a preference for pictograms and traditional bar charts. While pictograms enhance the narrative by adding an interesting visual element to the story, bar charts remain favored for their straightforward presentation and ease of understanding. Concerning the integration and modification of chart elements in narrative visualizations, we found that annotations and interactive elements could significantly boost the comprehensibility and engagement of visual representations. This suggests that a strategic combination of narrative elements with interactive data visualization can make complex information more accessible and engaging for users. Furthermore, pictograms used as annotations were identified as particularly effective in enhancing both comprehension and aesthetic appeal, suggesting that careful consideration of visual elements can improve the overall effectiveness of health communication tools. Yet, utilizing pictograms as bars and stacking pictograms is more memorable. Thus, the potential of these techniques for medical data stories is showcased.

For future work the exploration of different visualization techniques, such as age pyramids, is promising. They may offer clearer insights into demographic data distributions and trends over time. Visualization techniques that have not yet been considered can be investigated to broaden the DCs. The aim of this should be to create an extensive design space, which can aid in the creation of age visualizations for medical data stories. Additionally, integrating interactive elements could allow users to engage with the data, tailoring the information to their specific interests. Moreover, advanced technologies such as augmented and virtual reality could further improve the way we interact with data, offering immersive experiences that could significantly enhance the understanding and retention of complex information. The use of such innovative technologies, coupled with a strong narrative design, could lead to groundbreaking advancements in public health communication and education. 

\bibliographystyle{cag-num-names}
\bibliography{cag-template}

\begin{thebibliography}{60}
\providecommand{\natexlab}[1]{#1}
\providecommand{\url}[1]{\texttt{#1}}
\providecommand{\href}[2]{#2}
\providecommand{\path}[1]{#1}
\providecommand{\eprint}[1]{\href{http://arxiv.org/abs/#1}{\path{#1}}}
\providecommand{\DOIprefix}{doi:}
\providecommand{\ArXivprefix}{arXiv:}
\providecommand{\URLprefix}{URL: }
\providecommand{\Pubmedprefix}{pmid:}
\providecommand{\doi}[1]{\href{http://dx.doi.org/#1}{\path{#1}}}
\providecommand{\Pubmed}[1]{\href{pmid:#1}{\path{#1}}}
\providecommand{\BIBand}{and}
\providecommand{\bibinfo}[2]{#2}
\ifx\xfnm\undefined \def\xfnm[#1]{\unskip,\space#1}\fi
\bibitem[{{Robert Koch Institut (RKI)}(2022)}]{RKI_Hodenkrebs}
\bibinfo{author}{{Robert Koch Institut (RKI)}\xfnm[]}.
\newblock \bibinfo{title}{Hodenkrebs}.
\newblock
  \bibinfo{howpublished}{\url{https://www.krebsdaten.de/Krebs/DE/Content/Krebsarten/Hodenkrebs/hodenkrebs_node.html}};
  \bibinfo{year}{2022}.
\newblock \bibinfo{note}{Accessed: 2024-03-09}.
\bibitem[{Schaeffer et~al.(2021)Schaeffer, Berens, Vogt, Gille, Griese, Klinger
  et~al.}]{Schaeffer_2021}
\bibinfo{author}{Schaeffer\xfnm[ D]}, \bibinfo{author}{Berens\xfnm[ EM]},
  \bibinfo{author}{Vogt\xfnm[ D]}, \bibinfo{author}{Gille\xfnm[ S]},
  \bibinfo{author}{Griese\xfnm[ L]}, \bibinfo{author}{Klinger\xfnm[ J]}, et~al.
\newblock \bibinfo{title}{Health literacy in germany: Findings of a
  representative follow-up survey}.
\newblock \bibinfo{journal}{Deutsches Aerzteblatt International}
  \bibinfo{year}{2021};\bibinfo{volume}{118}(\bibinfo{number}{43}):\bibinfo{pages}{723}.
\bibitem[{Mittenentzwei et~al.(2022)Mittenentzwei, Weiß, Schreiber, Garrison,
  Bruckner, Pfister et~al.}]{Mittenentzwei_ArXive_2022}
\bibinfo{author}{Mittenentzwei\xfnm[ S]}, \bibinfo{author}{Weiß\xfnm[ V]},
  \bibinfo{author}{Schreiber\xfnm[ S]}, \bibinfo{author}{Garrison\xfnm[ LA]},
  \bibinfo{author}{Bruckner\xfnm[ S]}, \bibinfo{author}{Pfister\xfnm[ M]},
  et~al.
\newblock \bibinfo{title}{Narrative visualization to communicate neurological
  diseases}.
\newblock \bibinfo{year}{2022}.
\newblock \DOIprefix\doi{10.48550/ARXIV.2212.10121}.
\bibitem[{Meuschke et~al.(2022)Meuschke, Garrison, Smit, Bach, Mittenentzwei,
  Weiß et~al.}]{Meuschke_2022}
\bibinfo{author}{Meuschke\xfnm[ M]}, \bibinfo{author}{Garrison\xfnm[ LA]},
  \bibinfo{author}{Smit\xfnm[ NN]}, \bibinfo{author}{Bach\xfnm[ B]},
  \bibinfo{author}{Mittenentzwei\xfnm[ S]}, \bibinfo{author}{Weiß\xfnm[ V]},
  et~al.
\newblock \bibinfo{title}{Narrative medical visualization to communicate
  disease data}.
\newblock \bibinfo{journal}{Computers \& Graphics}
  \bibinfo{year}{2022};\bibinfo{volume}{107}:\bibinfo{pages}{144--157}.
\bibitem[{Segel and Heer(2010)}]{SegelandHeer_Narrative}
\bibinfo{author}{Segel\xfnm[ E]}, \bibinfo{author}{Heer\xfnm[ J]}.
\newblock \bibinfo{title}{Narrative visualization: Telling stories with data}.
\newblock \bibinfo{journal}{IEEE Transactions on Visualization and Computer
  Graphics}
  \bibinfo{year}{2010};\bibinfo{volume}{16}(\bibinfo{number}{6}):\bibinfo{pages}{1139--1148}.
\bibitem[{Bordegoni et~al.(2023)Bordegoni, Carulli and Spadoni}]{DouDiam}
\bibinfo{author}{Bordegoni\xfnm[ M]}, \bibinfo{author}{Carulli\xfnm[ M]},
  \bibinfo{author}{Spadoni\xfnm[ E]}.
\newblock \bibinfo{title}{User Experience and User Experience Design}.
\newblock \bibinfo{publisher}{Springer Nature Switzerland};
  \bibinfo{year}{2023}, p. \bibinfo{pages}{11–28}.
\newblock \DOIprefix\doi{10.1007/978-3-031-39683-0_2}.
\bibitem[{Kleinau et~al.(2022)Kleinau, Stupak, Mörth, Garrison, Mittenentzwei,
  Smit et~al.}]{Kleinau}
\bibinfo{author}{Kleinau\xfnm[ A]}, \bibinfo{author}{Stupak\xfnm[ E]},
  \bibinfo{author}{Mörth\xfnm[ E]}, \bibinfo{author}{Garrison\xfnm[ LA]},
  \bibinfo{author}{Mittenentzwei\xfnm[ S]}, \bibinfo{author}{Smit\xfnm[ NN]},
  et~al.
\newblock \bibinfo{title}{Is there a tornado in alex’s blood flow? a case
  study for narrative medical visualization}.
\newblock In: \bibinfo{booktitle}{Proc. of Eurographics Workshop on Visual
  Computing for Biology and Medicine}. \bibinfo{year}{2022}, p.
  \bibinfo{pages}{11--21}.
\bibitem[{B\"{o}ttinger et~al.(2020)B\"{o}ttinger, Kostis, Velez-Rojas,
  Rheingans and Ynnerman}]{Boettinger_2020}
\bibinfo{author}{B\"{o}ttinger\xfnm[ M]}, \bibinfo{author}{Kostis\xfnm[ HN]},
  \bibinfo{author}{Velez-Rojas\xfnm[ M]}, \bibinfo{author}{Rheingans\xfnm[ P]},
  \bibinfo{author}{Ynnerman\xfnm[ A]}.
\newblock \bibinfo{title}{Foundations of Data Visualization}; chap.
  \bibinfo{chapter}{Reflections on Visualization for Broad Audiences}.
\newblock \bibinfo{publisher}{Springer International Publishing};
  \bibinfo{year}{2020}, p. \bibinfo{pages}{297–305}.
\bibitem[{Winslow(1920)}]{Winslow}
\bibinfo{author}{Winslow\xfnm[ CEA]}.
\newblock \bibinfo{title}{The untilled fields of public health}.
\newblock \bibinfo{journal}{Science}
  \bibinfo{year}{1920};\bibinfo{volume}{51}(\bibinfo{number}{1306}):\bibinfo{pages}{23--33}.
\bibitem[{Campbell(2008)}]{Campbell2008hero}
\bibinfo{author}{Campbell\xfnm[ J]}.
\newblock \bibinfo{title}{The Hero with a Thousand Faces}.
\newblock Bollingen series; \bibinfo{publisher}{New World Library};
  \bibinfo{year}{2008}.
\bibitem[{Niccoli and Partridge(2012)}]{Niccoli2012}
\bibinfo{author}{Niccoli\xfnm[ T]}, \bibinfo{author}{Partridge\xfnm[ L]}.
\newblock \bibinfo{title}{Ageing as a risk factor for disease}.
\newblock \bibinfo{journal}{Current Biology}
  \bibinfo{year}{2012};\bibinfo{volume}{22}(\bibinfo{number}{17}):\bibinfo{pages}{R741–R752}.
\bibitem[{Hullman and Diakopoulos(2011)}]{Hullman2011}
\bibinfo{author}{Hullman\xfnm[ J]}, \bibinfo{author}{Diakopoulos\xfnm[ N]}.
\newblock \bibinfo{title}{Visualization rhetoric: Framing effects in narrative
  visualization}.
\newblock \bibinfo{journal}{IEEE Transactions on Visualization and Computer
  Graphics}
  \bibinfo{year}{2011};\bibinfo{volume}{17}(\bibinfo{number}{12}):\bibinfo{pages}{2231–2240}.
\bibitem[{Haroz et~al.(2015)Haroz, Kosara and Franconeri}]{Haroz_2015}
\bibinfo{author}{Haroz\xfnm[ S]}, \bibinfo{author}{Kosara\xfnm[ R]},
  \bibinfo{author}{Franconeri\xfnm[ SL]}.
\newblock \bibinfo{title}{{ISOTYPE} visualization: Working memory, performance,
  and engagement with pictographs}.
\newblock In: \bibinfo{booktitle}{Proc. of the ACM SIGCHI Conference on Human
  Factors in Computing Systems}. \bibinfo{year}{2015}, p.
  \bibinfo{pages}{1191--1200}.
\bibitem[{Bateman et~al.(2010)Bateman, Mandryk, Gutwin, Genest, McDine and
  Brooks}]{Bateman_2010}
\bibinfo{author}{Bateman\xfnm[ S]}, \bibinfo{author}{Mandryk\xfnm[ RL]},
  \bibinfo{author}{Gutwin\xfnm[ C]}, \bibinfo{author}{Genest\xfnm[ A]},
  \bibinfo{author}{McDine\xfnm[ D]}, \bibinfo{author}{Brooks\xfnm[ C]}.
\newblock \bibinfo{title}{Useful junk?: the effects of visual embellishment on
  comprehension and memorability of charts}.
\newblock In: \bibinfo{booktitle}{Proc. of the ACM SIGCHI Conference on Human
  Factors in Computing Systems}. \bibinfo{year}{2010}, p.
  \bibinfo{pages}{2573--2582}.
\bibitem[{Borkin et~al.(2013)Borkin, Vo, Bylinskii, Isola, Sunkavalli, Oliva
  et~al.}]{Borkin_2013}
\bibinfo{author}{Borkin\xfnm[ MA]}, \bibinfo{author}{Vo\xfnm[ AA]},
  \bibinfo{author}{Bylinskii\xfnm[ Z]}, \bibinfo{author}{Isola\xfnm[ P]},
  \bibinfo{author}{Sunkavalli\xfnm[ S]}, \bibinfo{author}{Oliva\xfnm[ A]},
  et~al.
\newblock \bibinfo{title}{What makes a visualization memorable?}
\newblock \bibinfo{journal}{IEEE Transactions on Visualization and Computer
  Graphics}
  \bibinfo{year}{2013};\bibinfo{volume}{19}(\bibinfo{number}{12}):\bibinfo{pages}{2306–2315}.
\bibitem[{Skau et~al.(2015)Skau, Harrison and Kosara}]{Skau_2015}
\bibinfo{author}{Skau\xfnm[ D]}, \bibinfo{author}{Harrison\xfnm[ L]},
  \bibinfo{author}{Kosara\xfnm[ R]}.
\newblock \bibinfo{title}{An evaluation of the impact of visual embellishments
  in bar charts}.
\newblock \bibinfo{journal}{Computer Graphics Forum}
  \bibinfo{year}{2015};\bibinfo{volume}{34}(\bibinfo{number}{3}):\bibinfo{pages}{221–230}.
\bibitem[{Li and Moacdieh(2014)}]{Li2014}
\bibinfo{author}{Li\xfnm[ H]}, \bibinfo{author}{Moacdieh\xfnm[ N]}.
\newblock \bibinfo{title}{Is “chart junk” useful? an extended examination
  of visual embellishment}.
\newblock \bibinfo{journal}{Proc of the Human Factors and Ergonomics Society
  Annual Meeting}
  \bibinfo{year}{2014};\bibinfo{volume}{58}(\bibinfo{number}{1}):\bibinfo{pages}{1516–1520}.
\bibitem[{Tufte(2001)}]{Tufte}
\bibinfo{author}{Tufte\xfnm[ ER]}.
\newblock \bibinfo{title}{The visual display of quantitative information};
  vol.~\bibinfo{volume}{2}.
\newblock \bibinfo{publisher}{Graphics press Cheshire, CT};
  \bibinfo{year}{2001}.
\bibitem[{Holmes(1984)}]{Holmes_1984}
\bibinfo{author}{Holmes\xfnm[ N]}.
\newblock \bibinfo{title}{Designer's Guide to Creating Charts \& Diagrams}.
\newblock \bibinfo{publisher}{Watson-Guptill Publications};
  \bibinfo{year}{1984}.
\bibitem[{Kim et~al.(2021)Kim, Setlur and Agrawala}]{Kim_2021}
\bibinfo{author}{Kim\xfnm[ DH]}, \bibinfo{author}{Setlur\xfnm[ V]},
  \bibinfo{author}{Agrawala\xfnm[ M]}.
\newblock \bibinfo{title}{Towards understanding how readers integrate charts
  and captions: A case study with line charts}.
\newblock In: \bibinfo{booktitle}{Proc. of the ACM SIGCHI Conference on Human
  Factors in Computing Systems}. \bibinfo{year}{2021}, p.
  \bibinfo{pages}{1--11}.
\bibitem[{Dimara and Perin(2020)}]{Dimara_2020}
\bibinfo{author}{Dimara\xfnm[ E]}, \bibinfo{author}{Perin\xfnm[ C]}.
\newblock \bibinfo{title}{What is interaction for data visualization?}
\newblock \bibinfo{journal}{IEEE Transactions on Visualization and Computer
  Graphics}
  \bibinfo{year}{2020};\bibinfo{volume}{26}(\bibinfo{number}{1}):\bibinfo{pages}{119–129}.
\bibitem[{Saket et~al.(2016)Saket, Endert and Stasko}]{Saket_2016}
\bibinfo{author}{Saket\xfnm[ B]}, \bibinfo{author}{Endert\xfnm[ A]},
  \bibinfo{author}{Stasko\xfnm[ J]}.
\newblock \bibinfo{title}{Beyond usability and performance: A review of user
  experience-focused evaluations in visualization}.
\newblock \bibinfo{address}{New York, NY, USA}: \bibinfo{publisher}{Association
  for Computing Machinery}; \bibinfo{year}{2016}, p.
  \bibinfo{pages}{133–142}.
\bibitem[{Borkin et~al.(2016)Borkin, Bylinskii, Kim, Bainbridge, Yeh, Borkin
  et~al.}]{Borkin2016}
\bibinfo{author}{Borkin\xfnm[ MA]}, \bibinfo{author}{Bylinskii\xfnm[ Z]},
  \bibinfo{author}{Kim\xfnm[ NW]}, \bibinfo{author}{Bainbridge\xfnm[ CM]},
  \bibinfo{author}{Yeh\xfnm[ CS]}, \bibinfo{author}{Borkin\xfnm[ D]}, et~al.
\newblock \bibinfo{title}{Beyond memorability: Visualization recognition and
  recall}.
\newblock \bibinfo{journal}{IEEE Transactions on Visualization and Computer
  Graphics}
  \bibinfo{year}{2016};\bibinfo{volume}{22}(\bibinfo{number}{1}):\bibinfo{pages}{519–528}.
\bibitem[{Kosara et~al.(2016)Kosara, Dasgupta and Bertini}]{Kosara_C4PGV_2016}
\bibinfo{author}{Kosara\xfnm[ R]}, \bibinfo{author}{Dasgupta\xfnm[ A]},
  \bibinfo{author}{Bertini\xfnm[ E]}.
\newblock \bibinfo{title}{Reflecting on the design criteria for explanatory
  visualizations}.
\newblock \bibinfo{journal}{Workshop on Creation, Curation, Critique and
  Conditioning of Principles and Guidelines in Visualization (C4PGV)}
  \bibinfo{year}{2016};.
\bibitem[{Attfield et~al.(2011)Attfield, Kazai, Lalmas and
  Piwowarski}]{Attfield_2011}
\bibinfo{author}{Attfield\xfnm[ S]}, \bibinfo{author}{Kazai\xfnm[ G]},
  \bibinfo{author}{Lalmas\xfnm[ M]}, \bibinfo{author}{Piwowarski\xfnm[ B]}.
\newblock \bibinfo{title}{Towards a science of user engagement (position
  paper)}.
\newblock In: \bibinfo{booktitle}{WSDM workshop on user Modelling for Web
  applications}; vol.~\bibinfo{volume}{1}. \bibinfo{organization}{Citeseer};
  \bibinfo{year}{2011},.
\bibitem[{Mahyar et~al.(2015)Mahyar, Kim and Kwon}]{mahyar2015}
\bibinfo{author}{Mahyar\xfnm[ N]}, \bibinfo{author}{Kim\xfnm[ SH]},
  \bibinfo{author}{Kwon\xfnm[ BC]}.
\newblock \bibinfo{title}{Towards a taxonomy for evaluating user engagement in
  information visualization}.
\newblock In: \bibinfo{booktitle}{Workshop on Personal Visualization: Exploring
  Everyday Life}; vol.~\bibinfo{volume}{3}. \bibinfo{year}{2015},
  p.~\bibinfo{pages}{4}.
\bibitem[{Cairo(2016)}]{cairo2016}
\bibinfo{author}{Cairo\xfnm[ A]}.
\newblock \bibinfo{title}{The truthful art: Data, charts, and maps for
  communication}.
\newblock \bibinfo{publisher}{New Riders}; \bibinfo{year}{2016}.
\bibitem[{Li(2020)}]{Li_2020}
\bibinfo{author}{Li\xfnm[ Q]}.
\newblock \bibinfo{title}{Overview of Data Visualization}.
\newblock \bibinfo{publisher}{Springer Singapore}.
\newblock ISBN \bibinfo{isbn}{9789811550690}; \bibinfo{year}{2020}, p.
  \bibinfo{pages}{17–47}.
\bibitem[{Moran(2017)}]{UsabilityAesthEffect}
\bibinfo{author}{Moran\xfnm[ K]}.
\newblock \bibinfo{title}{The aesthetic-usability effect}.
\newblock
  \bibinfo{howpublished}{\url{https://www.nngroup.com/articles/aesthetic-usability-effect/}};
  \bibinfo{year}{2017}.
\newblock \bibinfo{note}{Accessed: 2024-03-09}.
\bibitem[{Cawthon and Moere(2007)}]{Cawthon_2007}
\bibinfo{author}{Cawthon\xfnm[ N]}, \bibinfo{author}{Moere\xfnm[ AV]}.
\newblock \bibinfo{title}{The effect of aesthetic on the usability of data
  visualization}.
\newblock In: \bibinfo{booktitle}{Proc. of Information Visualization}.
  \bibinfo{year}{2007}, p. \bibinfo{pages}{637--648}.
\newblock \DOIprefix\doi{10.1109/IV.2007.147}.
\bibitem[{Moere and Purchase(2011)}]{Moere_2011}
\bibinfo{author}{Moere\xfnm[ AV]}, \bibinfo{author}{Purchase\xfnm[ H]}.
\newblock \bibinfo{title}{On the role of design in information visualization}.
\newblock \bibinfo{journal}{Information Visualization}
  \bibinfo{year}{2011};\bibinfo{volume}{10}(\bibinfo{number}{4}):\bibinfo{pages}{356–371}.
\bibitem[{Filonik and Baur(2009)}]{Filonik_2009}
\bibinfo{author}{Filonik\xfnm[ D]}, \bibinfo{author}{Baur\xfnm[ D]}.
\newblock \bibinfo{title}{Measuring aesthetics for information visualization}.
\newblock In: \bibinfo{booktitle}{2009 13th International Conference
  Information Visualisation}. \bibinfo{publisher}{IEEE}; \bibinfo{year}{2009},
  p. \bibinfo{pages}{579--584}.
\bibitem[{{World Health Organization (WHO)}(2024)}]{CommunNonCommunMental}
\bibinfo{author}{{World Health Organization (WHO)}\xfnm[]}.
\newblock \bibinfo{title}{Communicable and noncommunicable diseases, and mental
  health}.
\newblock
  \bibinfo{howpublished}{\url{https://www.who.int/our-work/communicable-and-non-communicable-diseases-and-mental-health}};
  \bibinfo{year}{2024}.
\newblock \bibinfo{note}{Accessed: 2024-03-09}.
\bibitem[{{World Health Organization (WHO)}(2022)}]{MentalDisorders}
\bibinfo{author}{{World Health Organization (WHO)}\xfnm[]}.
\newblock \bibinfo{title}{Mental disorders}.
\newblock
  \bibinfo{howpublished}{\url{https://www.who.int/news-room/fact-sheets/detail/mental-disorders}};
  \bibinfo{year}{2022}.
\newblock \bibinfo{note}{Accessed: 2024-03-09}.
\bibitem[{Sibbertsen and Lehne(2015)}]{Sibbertsen2015}
\bibinfo{author}{Sibbertsen\xfnm[ P]}, \bibinfo{author}{Lehne\xfnm[ H]}.
\newblock \bibinfo{title}{Eindimensionale empirische Verteilungen}.
\newblock \bibinfo{publisher}{Springer Berlin Heidelberg};
  \bibinfo{year}{2015}, p. \bibinfo{pages}{9–40}.
\bibitem[{{World Health Organization (WHO)}(2018)}]{Salmonella}
\bibinfo{author}{{World Health Organization (WHO)}\xfnm[]}.
\newblock \bibinfo{title}{Salmonella (non-typhoidal)}.
\newblock
  \bibinfo{howpublished}{\url{https://www.who.int/news-room/fact-sheets/detail/salmonella-(non-typhoidal)}};
  \bibinfo{year}{2018}.
\newblock \bibinfo{note}{Accessed: 2024-02-22}.
\bibitem[{{Bundesministerium für Gesundheit}(2022)}]{BPD}
\bibinfo{author}{{Bundesministerium für Gesundheit}\xfnm[]}.
\newblock \bibinfo{title}{Bipolare störung}.
\newblock
  \bibinfo{howpublished}{\url{https://gesund.bund.de/bipolare-stoerung}};
  \bibinfo{year}{2022}.
\newblock \bibinfo{note}{Accessed: 2024-02-22}.
\bibitem[{Hawke et~al.(2013)Hawke, Parikh and Michalak}]{Hawke2013}
\bibinfo{author}{Hawke\xfnm[ LD]}, \bibinfo{author}{Parikh\xfnm[ SV]},
  \bibinfo{author}{Michalak\xfnm[ EE]}.
\newblock \bibinfo{title}{Stigma and bipolar disorder: A review of the
  literature}.
\newblock \bibinfo{journal}{Journal of Affective Disorders}
  \bibinfo{year}{2013};\bibinfo{volume}{150}(\bibinfo{number}{2}):\bibinfo{pages}{181–191}.
\bibitem[{Manchia et~al.(2017)Manchia, Maina, Carpiniello, Pinna, Steardo,
  D’Ambrosio et~al.}]{Manchia2017}
\bibinfo{author}{Manchia\xfnm[ M]}, \bibinfo{author}{Maina\xfnm[ G]},
  \bibinfo{author}{Carpiniello\xfnm[ B]}, \bibinfo{author}{Pinna\xfnm[ F]},
  \bibinfo{author}{Steardo\xfnm[ L]}, \bibinfo{author}{D’Ambrosio\xfnm[ V]},
  et~al.
\newblock \bibinfo{title}{Clinical correlates of age at onset distribution in
  bipolar disorder: a comparison between diagnostic subgroups}.
\newblock \bibinfo{journal}{International Journal of Bipolar Disorders}
  \bibinfo{year}{2017};\bibinfo{volume}{5}(\bibinfo{number}{1}).
\bibitem[{Martin and Hanington(2013)}]{martin2013designmethoden}
\bibinfo{author}{Martin\xfnm[ B]}, \bibinfo{author}{Hanington\xfnm[ B]}.
\newblock \bibinfo{title}{Designmethoden: 100 Recherchemethoden und
  Analysetechniken f{\"u}r erfolgreiche Gestaltung}.
\newblock \bibinfo{publisher}{Stiebner Verlag GmbH}; \bibinfo{year}{2013}.
\bibitem[{Ren et~al.(2017)Ren, Brehmer, Lee, Höllerer and Choe}]{Ren2017}
\bibinfo{author}{Ren\xfnm[ D]}, \bibinfo{author}{Brehmer\xfnm[ M]},
  \bibinfo{author}{Lee\xfnm[ B]}, \bibinfo{author}{Höllerer\xfnm[ T]},
  \bibinfo{author}{Choe\xfnm[ EK]}.
\newblock \bibinfo{title}{Chartaccent: Annotation for data-driven
  storytelling}.
\newblock In: \bibinfo{booktitle}{Proc. of IEEE Pacific Visualization
  Symposium}. \bibinfo{year}{2017}, p. \bibinfo{pages}{230--239}.
\newblock \DOIprefix\doi{10.1109/PACIFICVIS.2017.8031599}.
\bibitem[{Wilke(2019)}]{wilke2019fundamentals}
\bibinfo{author}{Wilke\xfnm[ CO]}.
\newblock \bibinfo{title}{Fundamentals of data visualization: a primer on
  making informative and compelling figures}.
\newblock \bibinfo{publisher}{O'Reilly Media}; \bibinfo{year}{2019}.
\bibitem[{Tominski and Schumann(2020)}]{tominski2020interactive}
\bibinfo{author}{Tominski\xfnm[ C]}, \bibinfo{author}{Schumann\xfnm[ H]}.
\newblock \bibinfo{title}{Interactive visual data analysis}.
\newblock \bibinfo{publisher}{AK Peters/CRC Press}; \bibinfo{year}{2020}.
\bibitem[{Boy et~al.(2015)Boy, Detienne and Fekete}]{Boy2015}
\bibinfo{author}{Boy\xfnm[ J]}, \bibinfo{author}{Detienne\xfnm[ F]},
  \bibinfo{author}{Fekete\xfnm[ JD]}.
\newblock \bibinfo{title}{Storytelling in information visualizations: Does it
  engage users to explore data?}
\newblock In: \bibinfo{booktitle}{Proc. of ACM SIGCHI Conference on Human
  Factors in Computing Systems}. \bibinfo{address}{New York, NY, USA}:
  \bibinfo{publisher}{Association for Computing Machinery};
  \bibinfo{year}{2015}, p. \bibinfo{pages}{1449–1458}.
\bibitem[{Bach et~al.(2018)Bach, Stefaner, Boy, Drucker, Bartram, Wood
  et~al.}]{bach2018narrative}
\bibinfo{author}{Bach\xfnm[ B]}, \bibinfo{author}{Stefaner\xfnm[ M]},
  \bibinfo{author}{Boy\xfnm[ J]}, \bibinfo{author}{Drucker\xfnm[ S]},
  \bibinfo{author}{Bartram\xfnm[ L]}, \bibinfo{author}{Wood\xfnm[ J]}, et~al.
\newblock \bibinfo{title}{Narrative design patterns for data-driven
  storytelling}.
\newblock In: \bibinfo{booktitle}{Data-driven storytelling}.
  \bibinfo{publisher}{AK Peters/CRC Press}; \bibinfo{year}{2018}, p.
  \bibinfo{pages}{107--133}.
\bibitem[{Misoch(2019)}]{misoch2019qualitative}
\bibinfo{author}{Misoch\xfnm[ S]}.
\newblock \bibinfo{title}{Qualitative interviews}.
\newblock \bibinfo{publisher}{Walter de Gruyter GmbH \& Co KG};
  \bibinfo{year}{2019}.
\bibitem[{Feo(2023)}]{Dive_d}
\bibinfo{author}{Feo\xfnm[ E]}.
\newblock \bibinfo{title}{Inclusive design in 2024: Embracing diversity in
  every pixel}.
\newblock
  \bibinfo{howpublished}{\url{https://bootcamp.uxdesign.cc/inclusive-design-in-2024-embracing-diversity-in-every-pixel-f7a825ec1c14}};
  \bibinfo{year}{2023}.
\newblock \bibinfo{note}{Accessed: 2024-03-08}.
\bibitem[{Weir(2023)}]{Ageism}
\bibinfo{author}{Weir\xfnm[ K]}.
\newblock \bibinfo{title}{Ageism is one of the last socially acceptable
  prejudices. psychologists are working to change that}.
\newblock
  \bibinfo{howpublished}{\url{https://www.apa.org/monitor/2023/03/cover-new-concept-of-aging}};
  \bibinfo{year}{2023}.
\newblock \bibinfo{note}{Accessed: 2024-03-08}.
\bibitem[{Domínguez et~al.(2013)Domínguez, Alén and Fraiz}]{Domnguez2013}
\bibinfo{author}{Domínguez\xfnm[ T]}, \bibinfo{author}{Alén\xfnm[ E]},
  \bibinfo{author}{Fraiz\xfnm[ J]}.
\newblock \bibinfo{title}{International accessibility: a proposal for a system
  of symbols for people with disabilities}.
\newblock \bibinfo{journal}{International Journal on Disability and Human
  Development}
  \bibinfo{year}{2013};\bibinfo{volume}{12}(\bibinfo{number}{3}):\bibinfo{pages}{235--243}.
\newblock \DOIprefix\doi{10.1515/ijdhd-2012-0102}.
\bibitem[{Yudhanto et~al.(2022)Yudhanto, Pryhatyanto and
  Sulandari}]{Wireframing}
\bibinfo{author}{Yudhanto\xfnm[ Y]}, \bibinfo{author}{Pryhatyanto\xfnm[ WM]},
  \bibinfo{author}{Sulandari\xfnm[ W]}.
\newblock \bibinfo{title}{Designing and making ui/ux designs on the official
  website with the design thinking method}.
\newblock In: \bibinfo{booktitle}{2022 1st International Conference on Smart
  Technology, Applied Informatics, and Engineering (APICS)}.
  \bibinfo{year}{2022}, p. \bibinfo{pages}{165--170}.
\newblock \DOIprefix\doi{10.1109/APICS56469.2022.9918684}.
\bibitem[{Roberts(2011)}]{Roberts_2011}
\bibinfo{author}{Roberts\xfnm[ JC]}.
\newblock \bibinfo{title}{The five design-sheet (fds) approach for sketching
  information visualization designs.}
\newblock In: \bibinfo{booktitle}{Eurographics (Education Papers)}.
  \bibinfo{year}{2011}, p. \bibinfo{pages}{29--36}.
\bibitem[{Mannino and Abouzied(2019)}]{Mannino2019}
\bibinfo{author}{Mannino\xfnm[ M]}, \bibinfo{author}{Abouzied\xfnm[ A]}.
\newblock \bibinfo{title}{Is this real? generating synthetic data that looks
  real}.
\newblock In: \bibinfo{booktitle}{Proc. of the ACM Symposium on User Interface
  Software and Technology}. \bibinfo{address}{New York, NY, USA}:
  \bibinfo{publisher}{Association for Computing Machinery};
  \bibinfo{year}{2019}, p. \bibinfo{pages}{549–561}.
\bibitem[{{Institute for Health Metrics and Evaluation
  (IHME)}(2018)}]{WHOCollabIHME}
\bibinfo{author}{{Institute for Health Metrics and Evaluation (IHME)}\xfnm[]}.
\newblock \bibinfo{title}{Who and ihme collaborate to improve health data
  globally}.
\newblock
  \bibinfo{howpublished}{\url{https://www.healthdata.org/news-events/news-room/news-releases/who-and-ihme-collaborate-improve-health-data-globally}};
  \bibinfo{year}{2018}.
\newblock \bibinfo{note}{Accessed: 2024-03-03}.
\bibitem[{{Institute for Health Metrics and Evaluation (IHME)}(2023)}]{GBD}
\bibinfo{author}{{Institute for Health Metrics and Evaluation (IHME)}\xfnm[]}.
\newblock \bibinfo{title}{Global burden of disease (gbd)}.
\newblock
  \bibinfo{howpublished}{\url{https://www.healthdata.org/research-analysis/gbd}};
  \bibinfo{year}{2023}.
\newblock \bibinfo{note}{Accessed: 2024-03-03}.
\bibitem[{{Global Burden of Disease Collaborative
  Network}(2023)}]{GBD_Interactive}
\bibinfo{author}{{Global Burden of Disease Collaborative Network}\xfnm[]}.
\newblock \bibinfo{title}{Global burden of disease study 2019 (gbd 2019)}.
\newblock
  \bibinfo{howpublished}{\url{https://vizhub.healthdata.org/gbd-results/}};
  \bibinfo{year}{2023}.
\newblock \bibinfo{note}{Accessed: 2024-01-01}.
\bibitem[{Charters(2003)}]{ThinkingAloud}
\bibinfo{author}{Charters\xfnm[ E]}.
\newblock \bibinfo{title}{The use of think-aloud methods in qualitative
  research an introduction to think-aloud methods}.
\newblock \bibinfo{journal}{Brock Education Journal}
  \bibinfo{year}{2003};\bibinfo{volume}{12}(\bibinfo{number}{2}).
\newblock \DOIprefix\doi{10.26522/brocked.v12i2.38}.
\bibitem[{Tory and Moller(2005)}]{Tory2005}
\bibinfo{author}{Tory\xfnm[ M]}, \bibinfo{author}{Moller\xfnm[ T]}.
\newblock \bibinfo{title}{Evaluating visualizations: do expert reviews work?}
\newblock \bibinfo{journal}{IEEE Computer Graphics and Applications}
  \bibinfo{year}{2005};\bibinfo{volume}{25}(\bibinfo{number}{5}):\bibinfo{pages}{8--11}.
\newblock \DOIprefix\doi{10.1109/MCG.2005.102}.
\bibitem[{Garrison et~al.(2021)Garrison, Meuschke, Fairman, Smit, Preim and
  Bruckner}]{Garrison}
\bibinfo{author}{Garrison\xfnm[ L]}, \bibinfo{author}{Meuschke\xfnm[ M]},
  \bibinfo{author}{Fairman\xfnm[ J]}, \bibinfo{author}{Smit\xfnm[ NN]},
  \bibinfo{author}{Preim\xfnm[ B]}, \bibinfo{author}{Bruckner\xfnm[ S]}.
\newblock \bibinfo{title}{An exploration of practice and preferences for the
  visual communication of biomedical processes}.
\newblock \bibinfo{journal}{Eurographics Workshop on Visual Computing for
  Biology and Medicine} \bibinfo{year}{2021};:\bibinfo{pages}{1--12}.
\bibitem[{Brinch(2020)}]{Brinch}
\bibinfo{author}{Brinch\xfnm[ S]}.
\newblock \bibinfo{title}{16. What we talk about when we talk about beautiful
  data visualizations}.
\newblock \bibinfo{publisher}{Amsterdam University Press};
  \bibinfo{year}{2020}, p. \bibinfo{pages}{259–276}.
\bibitem[{G.~Aisch and Quealy(2015)}]{NYT}
\bibinfo{author}{G.~Aisch\xfnm[ AC]}, \bibinfo{author}{Quealy\xfnm[ K]}.
\newblock \bibinfo{title}{You draw it: How family income predicts children’s
  college chances}.
\newblock
  \bibinfo{howpublished}{\url{https://www.nytimes.com/interactive/2015/05/28/upshot/you-draw-it-how-family-income-affects-childrens-college-chances.html}};
  \bibinfo{year}{2015}.
\newblock \bibinfo{note}{Accessed: 2024-04-06}.

\end{thebibliography}



\end{document}